\documentclass[12pt, preprint]{aastex}

\def\etal{{\it et al.}~}
\def\kms{{km~s$^{-1}$}}

\def\arcmin{$^{\prime}~$}
\def\arcsec{$^{\prime\prime}~$}
\def\msun{$M_\odot$}

\def\h70i{{$h_{70}^{-1}$}}
\def\be{\begin{equation}}
\def\ee{\end{equation}}
\def\apjl{{ApJL}}
\def\apj{{ApJ}}
\def\apjs{{ApJS}}
\def\aj{{AJ}}
\def\aap{{A\&A}}
\def\aaps{{A\&AS}}

\def\mnras{{MNRAS}}
\def\nat{{Nature}}

\begin{document}

\title{Optically Unseen HI Detections towards the Virgo Cluster detected in the Arecibo Legacy Fast ALFA Survey}
\author {Brian R. Kent\altaffilmark{1}, Riccardo Giovanelli\altaffilmark{1,2}, Martha P. Haynes\altaffilmark{1,2}, 
Am\'elie Saintonge\altaffilmark{1},  Sabrina Stierwalt\altaffilmark{1}, Thomas Balonek\altaffilmark{2,3},
Noah Brosch\altaffilmark{4}, Barbara
Catinella\altaffilmark{5}, Rebecca A. Koopmann\altaffilmark{2,6},
Emmanuel Momjian\altaffilmark{5}, Kristine Spekkens\altaffilmark{7}}

\altaffiltext{1}{Center for Radiophysics and Space Research, Space Sciences Building,
Cornell University, Ithaca, NY 14853. {\it e--mail:} bkent@astro.cornell.edu, riccardo@astro.cornell.edu,
haynes@astro.cornell.edu, amelie@astro.cornell.edu, sabrina@astro.cornell.edu}

\altaffiltext{2}{National Astronomy and Ionosphere Center, Cornell University,
Space Sciences Building,
Ithaca, NY 14853. The National Astronomy and Ionosphere Center is operated
by Cornell University under a cooperative agreement with the National Science
Foundation.}

\altaffiltext{3}{Dept. of Physics \& Astronomy, Colgate University, Hamilton, NY 13346.
{\it e--mail:} tbalonek@mail.colgate.edu}

\altaffiltext{4}{The Wise Observatory \& The School of Physics and Astronomy,
Raymond \& Beverly Sackler Faculty of Exact Sciences, Tel Aviv University, Israel.
{\it e--mail:} noah@wise.tau.ac.il}

\altaffiltext{5}{Arecibo Observatory, National Astronomy and Ionosphere Center, 
Arecibo, PR 00612. {\it e--mail:} bcatinel@naic.edu, emomjian@naic.edu}

\altaffiltext{6}{Dept. of Physics \& Astronomy, Union College, Schenectady, NY 12308.
{\it e--mail:} koopmanr@union.edu}

\altaffiltext{7}{National Radio Astronomy Observatory and 
Dept. of Physics \& Astronomy, Rutgers University, 136 Frelinghuysen Road, Piscataway,
NJ 08854.  NRAO is a facility of the National Science Foundation operated under 
cooperative agreement by Associated Universities, Inc. {\it e--mail:} spekkens@physics.rutgers.edu}

\begin{abstract}

We report the discovery by the Arecibo Legacy Fast ALFA (ALFALFA)
survey of eight HI features not coincident with stellar counterparts
in the Virgo Cluster region. All of the HI clouds have $cz < $ 3000 \kms 
and, if at the Virgo distance, 
HI masses between 1.9$\times 10^7$ \msun~ and 1.1$\times 10^9$ \msun. Four of the
eight objects were reported or hinted at by previous studies and ``rediscovered'' 
by ALFALFA. While some clouds appear to be associated with optical galaxies in 
their vicinity, others show no clear association with a stellar counterpart.
Two of them are embedded 
in relatively dense regions of the cluster and are associated with M49 and M86; they 
were previously known. The others are mostly located in peripheral regions of the
cluster. Especially notable are a concentration of objects towards the so-called
M cloud, $3^\circ$ to $5^\circ$ to the NW of M87, and a complex of several clouds
projected roughly halfway between M87 and M49. The object referred to as VIRGOHI21
and proposed to be a ``dark galaxy''
is also detected and shown to be a tidal feature associated with NGC~4254.

\end{abstract}


\section{Introduction}

The predictions of the galaxy formation paradigm in the hierarchical scenario
requires corroboration by observational data.  Optically selected samples of
galaxies do not detect low luminosity, and hence presumably low mass objects,
in the predicted numbers. This has often been referred to as the ``substructure''
or ``missing satellite'' problem (Klypin \etal 1999). The possible existence of
a statistically significant population of dark matter dominated, optically faint 
halos would be of cosmological importance (Hawkins 1997; Somerville 2002).
While the statistics of halos that are completely devoid of baryons would be 
difficult to assess, optically faint but gas rich systems could be detected via 
their 21 cm line emission. ALFALFA, the Arecibo Legacy Fast ALFA extragalactic 
HI survey (Giovanelli \etal 2005) currently underway, will cover 7000 square degrees 
of sky at $cz <$ 18000 \kms.  At this time more than half of the solid angle 
encompassing the Virgo Cluster has been fully surveyed.  ALFALFA can detect 
$\sim 2\times 10^7$ \msun~ at the cluster distance, which in this paper will 
be assumed to be 16.7 Mpc.  The Virgo Cluster offers a fertile environment for 
the possible detection of gas-rich, optically faint systems.

A number of extragalactic HI clouds have been reported in the past 
(e.g. Schneider \etal 1983; Sancisi \etal 1987; Giovanelli \& Haynes 1989; 
Chengalur \etal 1995; Kilborn \etal 2000; Ryder \etal 2001; Minchin \etal 2005a; 
Oosterloo \& van Gorkom 2005); they are however not necessarily identified with 
optically dark halos. Some have been shown to be tidal appendages or to be associated with 
optical counterparts, e.g. the NE component of the HI~1225+01 pair of objects
(Giovanelli, Williams \& Haynes 1991) and the object known as VIRGOHI21 (Davies \etal ~2004;
Minchin \etal ~2005a,b; Haynes \etal ~in preparation); others are thought
to be the result of ram pressure stripping in the cluster environment, e.g. the Oosterloo \& 
van Gorkom feature near NGC 4388. The interactions between sub-groups in Virgo 
play an important role in the cluster's evolution, acting as a preprocessing step
of the material as the galaxies fall into the cluster.

Here we present a catalog of 21 cm line sources detected by ALFALFA in the central portion
of the Virgo Cluster region which have no obvious optical counterparts.  
Parameters of the detections and descriptions of their environments are given. Three 
of the sources were previously reported and one hinted at by other studies and 
``rediscovered'' by ALFALFA.

\section{Observations and Source Parameters}

The sources presented in this paper are part of the ALFALFA catalog and refer to 
the region $12^h <$ R.A.(J2000) $< 13^h$ and $+8^\circ <$ Dec.(J2000) $<+16^\circ$.
Complete source catalogs of this region have been reported by Giovanelli \etal (2007) 
and Kent \etal (2007 in preparation); they are accessible at 
http://arecibo.tc.cornell.edu/hiarchive/alfalfa/. Sources are extracted from the
ALFALFA data set via an automated algorithm (Saintonge 2007), successively
inspected by eye, measured and classified according to a code which primarily 
depends on signal to noise (S/N). The objects reported in this paper are all 
classified as {\it bona fide} detections (code 1; see Giovanelli \etal 2007), 
typically of S/N $\gtrsim$ 6.5, where S/N is defined as 
\be
        S/N=\Bigl({1000 F_c\over W_{50}}\Bigr){w^{1/2}_{smo}\over \sigma_{rms}}
\ee
        where $F_c$ is the integrated flux density in Jy \kms,
        $w_{smo}$ is either $W_{50}/(2\times 10)$ for $W_{50}<400$ \kms ~or
        $400/(2\times 10)=20$ for $W_{50} \geq 400$ \kms ~[$w_{smo}$ is a
        smoothing width expressed as the number of spectral resolution
        bins of 10 \kms ~bridging half of the signal width], and $\sigma_{rms}$
        is the r.m.s noise figure across the spectrum measured in mJy at 10
        \kms ~resolution.
All of them have been confirmed by corroborating observations carried out with
the Arecibo telescope or with the Very Large Array (VLA) as discussed below. 
Details of the ALFALFA observations can be found in Giovanelli \etal (2005)
and Giovanelli \etal (2007).

Table 1 contains the observed and derived parameters of the HI clouds. Their 
velocities indicate that an association with the Virgo cluster is possible 
for most of them, with the exception of the group associated with NGC 4795/4796. 
The latter is more likely to be in the background of the Virgo cluster, at a 
distance of $\sim$40 Mpc which is assumed for the clouds in that group. 

The fields of each of the sources in Table 1 have been inspected in 
the Sloan Digital Sky Survey\footnote{
Funding for the Sloan Digital Sky Survey (SDSS) has been 
provided by the Alfred P. Sloan Foundation, the Participating Institutions, 
the National Aeronautics and Space Administration, the National Science 
Foundation, the U.S. Department of Energy, the Japanese Monbukagakusho, 
and the Max Planck Society. The SDSS Web site is http://www.sdss.org/.
The SDSS is managed by the Astrophysical Research Consortium (ARC) for 
the Participating Institutions. The Participating Institutions are The 
University of Chicago, Fermilab, the Institute for Advanced Study, the 
Japan Participation Group, The Johns Hopkins University, Los Alamos National 
Laboratory, the Max-Planck-Institute for Astronomy (MPIA), the Max-Planck-Institute 
for Astrophysics (MPA), New Mexico State University, University of Pittsburgh, 
Princeton University, the United States Naval Observatory, and the University 
of Washington.}
and the DSS2 via {\it Skyview.}\footnote{{\it Skyview} was developed and 
maintained under NASA ADP Grant NAS5--32068 under the auspices of the High 
Energy Astrophysics Science Archive Research Center at the Goddard Space 
Flight Center Laboratory of NASA.}  The contents of Table 1 are as follows.\\
\textit{Col.(1)} - Cloud ID number\\
\textit{Col.(2 \& 3)} - HI source center coordinates (J2000); these positions are typically
accurate to within 30\arcsec or better (see Giovanelli \etal 2007)\\
\textit{Col.(4)} - Heliocentric velocity in \kms\\
\textit{Col.(5)} - Velocity width measured at half peak power in \kms\\
\textit{Col.(6)} - Integrated flux in Jy \kms\\
\textit{Col.(7)} - Signal to noise ratio\\
\textit{Col.(8)} - Base 10 logarithm of the HI mass in solar units, assuming HI is optically thin\\
\textit{Col.(9)} - Angular distance from M87 in degrees\\  

For three of the sources, we separately list the parameters of several clumps,
identified with italic qualifiers.
Figure 1 shows locations of the sources within the Virgo cluster region; the 
grayscale background image shows hard X-ray counts (0.5-2.0 keV) from the ROSAT 
dataset of Snowden \etal (1995), smoothed with a 5\arcmin ~kernel.  The approximate
boundaries of the M and W\arcmin ~clouds are indicated by dashed circles (Binggeli, 
Popescu \& Tammann 1993). 
We note that the fields of the HI clouds often contain one or several small
optical objects; the possibility that one of them may be a small dwarf 
or low surface brightness galaxy associated with the HI source cannot be excluded at this time.
We discuss the characteristics of each HI source below.

\textit{Cloud 1}.--This object, unresolved by the Arecibo beam of  3.3\arcmin $\times$ 
3.8\arcmin,
is near the detection limit of ALFALFA at the Virgo distance; S/N and spectrum used are
those of the ALFALFA survey observations. That detection has been confirmed by 
successive, more sensitive Arecibo observations. This is the object with the lowest
HI mass in Table 1. We note that the Irr galaxy UGC~7003
lies 30\arcmin ~NW of the HI source at $cz= 1286$ \kms, the NGC~4019 group lies 15\arcmin 
~W at $cz= 1524$ \kms~ and UGC~7038 lies 32\arcmin NW at $cz= 889$ \kms.

\textit{Cloud 2}.--This object, also unresolved by the Arecibo beam, lies 3.8\arcmin 
away from a small optical galaxy for which no optical redshift is known.
The optical galaxy AGC~220171 at 121035.6+114539, an apparently undisturbed object classified
as a BCD, lies 29\arcmin ~to the SE at a redshift of $cz= 1296$ \kms.  A VLA map 
of the source has been obtained and the results will be discussed in a forthcoming study.

\textit{Cloud 3}.--This object, also unresolved by the Arecibo beam, is located
in a crowded field - a host of galaxies are in the surrounding periphery at 
a comparable redshift.  The density of galaxies with known optical or HI redshifts 
in this region is high and 21 are known with $1800 < cz < 2500$ \kms~ within 
one degree of the HI source. The nearest optical galaxy with similar velocity is 
AGC 221651 at 121336.4+130201, 8\arcmin N of the HI feature at $cz= 1932$ \kms.
A VLA map of the source has  been obtained and the results will be discussed 
in a forthcoming study.
 
\textit{Cloud 4}.--The HI source is located 3.6\arcmin southeast of M86 (=NGC 4406; $cz = -244$ \kms), the 
large Virgo S0 galaxy. It was previously reported by Davies 
\etal (2004) as VIRGOHI4. They also suggested that the source could lie behind 
M86. HI synthesis imaging by Oosterloo \& van Gorkom (2005)
showed the HI source to be a plume extending from NGC~4388($cz=$2524 \kms) and possibly resulting
from interaction between that galaxy and the intracluster gas.  Jacoby \etal (2005) indicated
that H$\alpha$ filaments detected could be associated with this plume.
In the same vicinity ALFALFA makes a separate detection of M86 (Giovanelli \etal 2007) at
negative velocities with the HI centroid located 18$^{\prime\prime}$ to 
the south.
The features reported by Bregman \& Roberts (1990) are associated with the 
ALFALFA HI detection of
M86.

\textit{Cloud 5}.--This object was first reported by Sancisi \etal (1987). Later, 
aperture synthesis observations were presented by Henning \etal (1993) and 
McNamara \etal (1994). The HI source is located 2.6\arcmin southeast of M49($cz= 997$ \kms) and 
it is proposed to be related to the interaction between M49 and the dwarf 
irregular UGC~7636($cz= 276$ \kms).

\textit{Cloud 6}.--Two sources(6{\it a\&b}) are the brightest clumps in a stream found
in the vicinity of NGC~4254(M99; $cz= 2407$ \kms). The brightest of the two is an object first detected
at Jodrell Bank (Davies \etal ~2004; Minchin \etal 2005a)
and more recently mapped at Westerbork by the same team (Minchin 
\etal 2005b).  Dubbed ``VirgoHI21'', it was proposed by that group that the source
is a giant ``dark galaxy.'' The ALFALFA observations clearly show the feature to
be part of a stream connected to NGC~4254, continuously extending some 250 kpc to 
the North of that galaxy. The feature appears of tidal origin. 
 A third ALFALFA detection
(6{\it c}) lies 18 \arcmin west of NGC~4254.  A more detailed 
analysis of the ALFALFA evidence on this object is given in Haynes \etal (2007).

\textit{Cloud 7}.--A complex of five HI clouds, it projects between
M87 and M49, roughly $3^\circ$ south of the former. The five clouds,
spread in velocity between 480 and 607 \kms, extend over approximately
35\arcmin in the plane of the sky, or 200 kpc at the Virgo cluster distance.
In HI mass, the clouds range between $0.5\times 10^8$ and $2.0\times 10^8$
\msun. VLA observations have been obtained and clouds 7c and 7d have
been clearly detected. Detailed results of both the ALFALFA and VLA
observations are in preparation by Kent {\it et al.}

\textit{Cloud 8}.--Three clouds comprise an HI complex that surrounds 
the SB0/a galaxy NGC~4795($cz=2781$ \kms) and its dwarf companion NGC~4796($cz=2406$ \kms). It was previously
noted that a large flux measurement discrepancy between Arecibo and Effelsberg 
measurements was likely due to an offset of the HI from the center of the 
NGC~4795/4796 pair (Hoffman \etal 1989). Later observations at Arecibo by
Duprie \& Schneider (1996) are suggestive of an extension of the HI
emission from NGC~4796 towards its irregular companions UGC~8042/5,
perhaps arising from a tidal interaction. ALFALFA maps show that HI emission
surrounds the NGC~4795/6 pair, and that it is indeed connected to the two 
nearby galaxies, UGC~8045 and UGC~8042, which are respectively at 
$cz= 2801$ and $cz= 2856$ \kms.  The HI masses computed in Table 1
assume a distance of $\sim$40 Mpc. The kinematics of the HI indicate 
that all the galaxies in this system are at a comparable redshift and are 
interacting as a group.

\section{Discussion}

The characteristics of the eight sources reported here are quite diverse.
They can be grouped into three categories:  isolated objects, objects in 
the vicinity of large galaxies, and disturbed objects that could be remnants 
of an encounter with a larger system, such as the collective cluster potential.  
Except for cloud 8, the projected distances to M87 and heliocentric velocities 
place the HI sources within the canonical boundaries of the Virgo Cluster, as 
defined by Binggeli, Sandage, \& Tammann (1985).

Clouds 4 and 5 have been extensively studied in the past, using higher
aperture synthesis data which indicate a clear association with Virgo
cluster galaxies M49 and NGC~4388. The proximity to these galaxies and the
cluster environment are important in determining their properties.
The clouds in the vicinity of NGC~4795 (cloud 8) are most likely the
result of tidal stripping in interacting members of a close group, not 
an uncommon occurrence. 

Cloud 6 has raised significant attention, as it
was proposed to be a rare representative of a category of massive, yet
starless galaxies. The ALFALFA observations clearly show that rather
than an isolated ``dark galaxy,'' this object is an extended stream
connected to NGC~4254. While no nearby companion with which NGC~4254
may have had a close encounter is clearly identifiable, the origin of
the feature appears most likely to be of a tidal nature, an event of
``harassment'' of the type graphically illustrated in simulations, e.g. 
by Moore \etal (1996) and Lake \etal (1998). 

ALFALFA observations of clouds 1, 2 and 3 offer no clear hints as to 
their origin. They are all unresolved by the Arecibo beam, hence their
HI is contained within a $\sim 10$ kpc diameter or less. The HI masses 
of the clouds are relatively low, especially for clouds 1 and 2, and 
the lack of a size determination impedes an estimate of their dynamical masses; 
upper limits for the latter on order of 1--6$\times 10^8$ \msun ~can be 
obtained if one assumes turbulent or rotational motion amplitudes as indicated 
by the velocity widths. These objects are projected near or within the boundaries of 
the M cloud, a loose subclump thought to be behind and falling into 
the main cluster around M87, although little coherence in the velocities
of the trio exists to firmly substantiate such association. At any rate,
they appear to be relatively isolated and far removed from the central parts 
of the cluster so that gravitational, rather than hydrodynamical processes 
involving the intracluster gas, are more likely to be invoked in explaining 
their nature. While the mean number density of VCC galaxies in the M cloud 
region is less than 1.5 $\times 10^{-3}$ (arcmin)$^{-2}$ (Binggeli, Sandage, \& Tammann 1985, Schindler 
\etal 1999), cloud 3 lies in a locally overdense region, where galaxy-galaxy 
interactions may be more frequent. It cannot be excluded however that these  
clouds may be primordial, low mass halos, perhaps associated with
small dwarf or low surface brightness
optical counterparts. Analysis of recently obtained   
VLA data and planned follow-up optical studies will help elucidate this issue.

Cloud 7 is resolved by ALFALFA data into several separate clouds,
spread over more than 200 kpc (if located at the cluster distance)
and 250 \kms ~in extent. With a mean velocity near 540 \kms, the complex
could well be located in the foreground of the cluster, although cluster
galaxies of similar redshift are found in the vicinity. Most notably,
NGC~4424, an SBa at $cz=476$ \kms ~is located about 40\arcmin ~to the west
of the complex. ALFALFA and VLA HI maps of that object (Chung \etal 2007),
as well as CO maps (Cortes \etal 2006) 
indicate that its structure is disturbed, showing an appendage extending
to the south of the galaxy, pointing opposite
the direction of the cloud complex. Assuming the whole of the complex is not a
gravitationally bound unit, the velocity differences between the individual
clouds will separate them at the approximate rate of $\sim 250$ kpc Gyr$^{-1}$.
Differential motions of this amplitude are consistent with tidal forces
within the cluster potential and suggest the complex may disperse within
a cluster crossing time. A more detailed analysis based on ALFALFA and
VLA observations will be explored in a future study (Kent {\it et al.}, in preparation).

While a detailed statistical study of the ALFALFA catalogs in the Virgo
regions awaits completion of the survey effort therein, it is interesting
to preliminarily note that ALFALFA does not detect very large numbers of low 
HI mass sources in the cluster. This is an indication that, at least in the
cluster region, the HI mass function faint end slope does not rise 
sufficiently to significantly contribute to solving the ``substructure''
problem mentioned in our introduction.

We gratefully acknowledge Dr. Tom Oosterloo
for his critical reading of this paper and
providing comments.  This research has made use of the NASA/IPAC Extragalactic Database (NED) which is 
operated by the Jet Propulsion Laboratory, California Institute of Technology, 
under contract with the National Aeronautics and Space Administration.  KS is 
a Jansky Fellow. 
This work has been supported by NSF 
grants AST--0307661, AST--0435697, AST--0607007 and the Brinson Foundation.

\begin{deluxetable}{rrrrcrrcl}
\tablecaption{ALFALFA Optically Unseen Detections \label{AOparams}}
\tablewidth{0pt}
\tabletypesize{\footnotesize}
\tablehead{
              \colhead{Cloud ID} 
	    & \colhead{$\alpha$} 
	    & \colhead{$\delta$} 
	    & \colhead{$cz_\odot$} 
	    & \colhead{$W_{\rm{FWHM}}$} 
	    & \colhead{$F_{c}$} 
	    & \colhead{S/N} 
	    & \colhead{log$_{10} M_{HI}$}   
	    & \colhead{d$_{M87}$} \\
	    \colhead{} 
	    & \colhead{J2000} 
	    & \colhead{J2000} 
	    & \colhead{(\kms)} 
	    & \colhead{(\kms)} 
	    & \colhead{(Jy \kms)}  
	    & \colhead{} 
	    & \colhead{\msun} 
	    & \colhead{deg}
         }

\startdata

 1\tablenotemark{a} & 12 02 44.4 & +14 04 56  & 1121$\pm$  1  &   22$\pm$  2  &   0.30 $\pm$ 0.02 &   5.1   & 7.29 & 7.0\\
 2\tablenotemark{b} & 12 08 45.5 & +11 55 17  & 1230$\pm$  1  &   29$\pm$  2  &   0.77 $\pm$ 0.04  & 11.6   & 7.63 & 5.4\\
 3\tablenotemark{b} & 12 13 41.8 & +12 53 51  & 2235$\pm$  2  &   53$\pm$  3  &   1.21 $\pm$ 0.07  &  9.2   & 8.54 & 4.2 \\
 4\tablenotemark{c} & 12 26 19.4 & +12 53 30  & 2246$\pm$  5  &  135$\pm$ 11  &   2.05 $\pm$ 0.07  & 14.4   & 8.77 & 1.2 \\
 5\tablenotemark{d} & 12 29 54.7 & +07 58 12  &  473$\pm$  5  &   30$\pm$ 10  &   1.13 $\pm$ 0.06  & 10.9   & 7.87 & 4.4 \\
 6{\it a}\tablenotemark{e} & 12 17 55.5 & +14 44 45  & 1984$\pm$  1  &  128$\pm$  2  &   2.09 $\pm$ 0.06  & 16.2   & 8.70 & 3.9 \\
 6{\it b}\tablenotemark{e} & 12 17 49.1 & +15 04 52  & 2200$\pm$  6  &   40$\pm$ 13  &   0.52 $\pm$ 0.05  &  5.0   & 8.16 & 4.1 \\
 6{\it c}\tablenotemark{e} & 12 17 33.8 & +14 23 47  & 2111$\pm$  10 &   65$\pm$ 20  &   0.57 $\pm$ 0.04  &  7.3   & 8.17 & 3.8 \\
7{\it a}\tablenotemark{b} & 12 29 42.8 & +09 41 54  &  524$\pm$  7  &  116$\pm$ 15  &   1.16 $\pm$ 0.07  &  8.6   & 7.87 & 2.7   \\
7{\it b}\tablenotemark{b} & 12 30 19.4 & +09 35 18  &  603$\pm$  4  &  252$\pm$  7  &   2.56 $\pm$ 0.09  & 13.1   & 8.22 & 2.8 \\
7{\it c}\tablenotemark{b} & 12 30 25.8 & +09 28 01  &  488$\pm$  5  &   62$\pm$ 11  &   2.48 $\pm$ 0.07  & 21.2   & 8.21 & 2.9 \\
7{\it d}\tablenotemark{b} & 12 31 19.0 & +09 27 49  &  607$\pm$  4  &   56$\pm$  7  &   0.72 $\pm$ 0.06  &  6.5   & 7.67 & 2.9 \\
7{\it e}\tablenotemark{b} & 12 31 26.7 & +09 18 52  &  480$\pm$ 10  &   53$\pm$ 21  &   0.91 $\pm$ 0.06  &  7.6   & 7.77 & 3.1 \\
8{\it a}\tablenotemark{f} & 12 55 04.3 & +08 06 13  & 2629$\pm$  3  &   71$\pm$  7  &   0.72 $\pm$ 0.07  &  6.4   & 8.43 & 7.3 \\
8{\it b}\tablenotemark{f} & 12 55 10.2 & +08 02 44  & 2754$\pm$ 14  &  407$\pm$ 27  &   2.91 $\pm$ 0.12  &  9.4   & 9.08 & 7.4 \\
8{\it c}\tablenotemark{f} & 12 55 13.7 & +08 02 51  & 2771$\pm$  4  &  292$\pm$  7  &   2.52 $\pm$ 0.10  & 10.9   & 9.02 & 7.4 \\

\enddata
\tablenotetext{a}{Confirmed by follow-up, high sensitivity observation at Arecibo.}
\tablenotetext{b}{VLA maps obtained, processing underway.}
\tablenotetext{c}{VirgoHI4 (Davies \etal 2004), VLA map by Oosterloo \etal (2005).}
\tablenotetext{d}{Vicinity of M49, Sancisi \etal (1987); synthesis data by Henning \etal (1993).}
\tablenotetext{e}{Clumps in VirgoHI21, WSRT data by Minchin \etal (2005).}
\tablenotetext{f}{NGC 4795/4796 group.}
\end{deluxetable}

\clearpage

\begin{figure}
\plotone{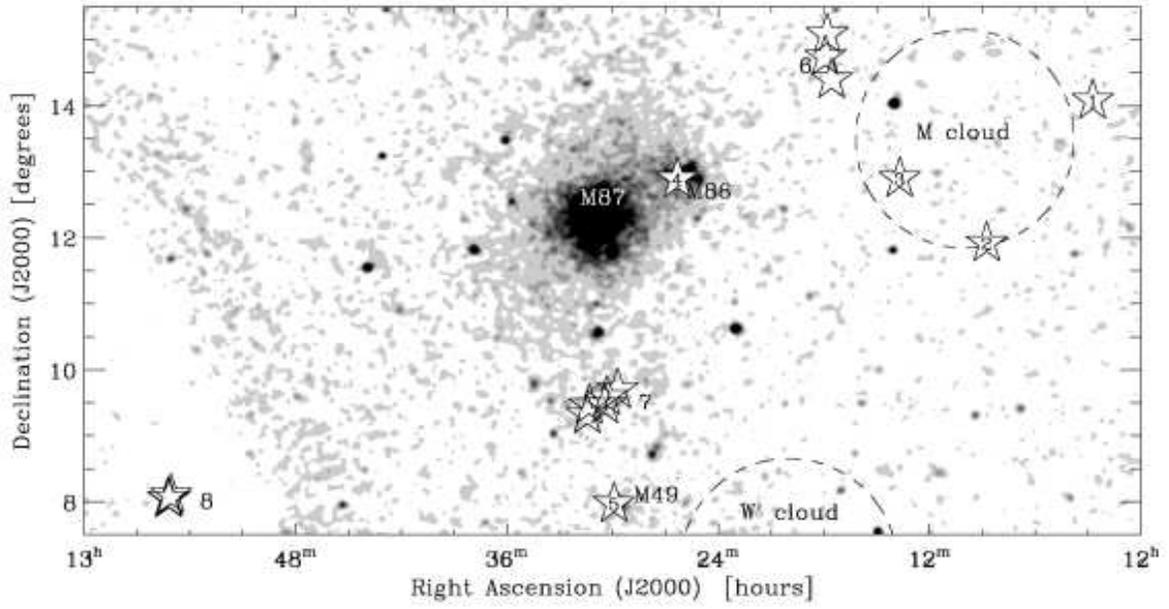}
\caption{Sky distribution of the HI detections (star symbols and corresponding Table 1 numbers) presented in this paper.
The X-ray peaks of cluster members M87, M86, and M49 are labeled for reference.  The background grayscale image is a
hard X-ray counts image from ROSAT (Snowden \etal 1995), smoothed with a 5\arcmin Gaussian kernel.  
The portions of the M and W\arcmin~ subclouds are indicated by the dashed circles (Binggeli, Popescu, \& Tammann 1993).}
\end{figure}

\end{document}